\documentclass[a4paper]{panl}
\usepackage{cite}
\usepackage{wrapfig}
\usepackage{graphicx}
\usepackage{amssymb}
\usepackage{amsfonts}
\usepackage{amsmath}
\usepackage{longtable}
\usepackage{rotating}
\usepackage{lscape}
\usepackage{epsfig}
\usepackage{multirow}

\originalTeX
\begin{document}

\title{Multimessenger astronomy  \\ Многоканальная астрономия}
\maketitle

\authors{V.\,Rozhkov\,$^{a,}$\footnote{E-mail: rozhkov@jinr.ru},
S.\,Troitsky\,$^{b,c,}$\footnote{E-mail: st@ms2.inr.ac.ru}}

\setcounter{footnote}{0}

\authors{В.\,Рожков\,$^{a,1}$, С.\,Троицкий\,$^{b,c,2}$}

\from{$^{a}$\,International Intergovernmental Scientific Research Organization Joint Institute for Nuclear Research, 6 Joliot-Curie St., 141980, Dubna, Moscow Region, Russia}
\from{$^{a}$\,Международная межправительственная научно-исследовательская организация Объединённый институт ядерных исследований, ул. Жолио-Кюри, 6, 141980, г. Дубна, Московская обл., Россия}
\from{$^{b}$\,Institute for Nuclear Research of the Russian Academy of Sciences, 60th October Anniversary prospect~7a, 117312 Moscow, Russia}
\from{$^{b}$\,Институт ядерных исследований Российской академии наук, проспект 60-летия Октября, 7А, 117312 Москва, Россия}
\from{$^{c}$\,Physics Department, Lomonosov Moscow State University, 1-2 Leninskie Gory, 119991 Moscow, Russia}
\from{$^{c}$\,Физический факультет Московского государственного университета имени М.В. Ломоносова, Ленинские горы, 1-2, 119991 Москва, Россия}

\begin{abstract}
Этот краткий обзор составлен на основе лекции, прочитанной одним из авторов на международной молодежной конференции AYSS-2023, и посвящен многоканальной астрономии, изучающей астрофизические объекты и явления с использованием частиц и волн различных типов, приносящих информацию из космоса. Они включают электромагнитные и гравитационные волны, нейтрино и космические лучи. Обсуждаются новые возможности, открывающиеся благодаря совместному использованию разных носителей информации. Сочетание результатов, полученных с помощью различных каналов наблюдения, позволяет получать более полную и точную информацию о процессах, происходящих во Вселенной, и даже использовать ее для изучения фундаментальной физики.
\vspace{0.2cm}

This brief review is based on a lecture given by one of the authors at the international youth conference AYSS-2023. It is devoted to multimessenger astronomy, which studies astrophysical objects and phenomena using various particles and waves that bring information from space. The messengers include electromagnetic and gravitational waves, neutrinos, and cosmic rays. We discuss new opportunities that open up with the combined use of several carriers of information. Combination of data obtained through various observation channels allows one to obtain more complete and accurate information about the processes occurring in the Universe, and even to use it for studying fundamental physics.
\end{abstract}
\vspace*{6pt}

\noindent


\label{sec:intro}
\section{Introduction}
Modern astronomy makes its best out of technological developments and space exploration, which open up novel observational techniques. One of the most promising and dynamically developing branch is the multimessenger astronomy, an approach that uses not only traditional electromagnetic waves, but also other astrophysical messengers such as neutrinos, cosmic rays and gravitational waves.

Our brief review of multimessenger astronomy aims to provide readers with a general understanding of this exciting area of research. We briefly cover a wide range of topics, ranging from the history and development of multimessenger astronomy to the latest advances and discoveries made within this approach. The purpose of this lecture is to demonstrate how multimessenger astronomy changes our vision of the Universe, and how it contributes to the development of fundamental physics. We discuss the key cosmic messengers, including photons of various bands of electromagnetic radiation, neutrinos, gravitational waves, and high-energy cosmic rays. We highlight the benefits of working with such a diverse set of data simultaneously, and present examples of application of this approach. The review is aimed at the newcomers to the field as well as at anyone interested, but some more experienced scientists might also find parts of it useful. Many educational resources, including a variety of books and lectures, aim to make this complex field accessible to both specialists and the general public. Because of the introductory nature of this lecture, we generally do not give references to original scientific publications, limiting the bibliography to books and review articles only; all original references may be found there. Those interested to go beyond the present  lecture may start e.g.\ with excellent books \cite{Cherepaschuk}\cite{Capone}\cite{Spurio-book}. 

The strength of the multimessenger astronomy is in its integrative approach. By combining observations from different messengers, one can construct a more complete and nuanced model of the Universe's most dramatic and energetic events. This method enhances not only the depth of data but also the precision of the interpretations, offering insights that surpass what could be achieved using any single type of data alone.

Historically tied to the fundamental physics, this field leverages the Universe as a vast laboratory, exploring  fundamental forces and particles that compose the cosmic landscape. As such, the multimessenger astronomy is pivotal in pushing the boundaries of our understanding of the Universe both at macroscopic and microscopic scales. 

The rest of the paper is organized as follows. In Sec.~\ref{sec:messengers}, we list and briefly discuss four kinds of cosmic messengers, that is photons, neutrinos, gravitational waves and cosmic rays. Section~\ref{sec:examples} presents two canonical examples of multimessenger observations of single burst-like events. In Sec.~\ref{sec:diffuse}, we address the multimessenger approach in the case of integrated diffuse radiation. We briefly conclude in Sec.~\ref{sec:Conclusion}.

~

\section{Photons and beyond: astrophysical messengers}
\label{sec:messengers}
Multi-messenger astronomy generalizes the traditional astronomy, based on detection of the electromagnetic radiation, that is photons, from cosmic objects and events. Within this approach, conventional astronomical observations are supplemented by the detection of neutrinos, gravitational waves, and a diverse group of particles known as cosmic rays, which encompasses protons, nuclei, electrons and positrons and even some exotic hypothetical particles. Neutrinos, with their minimal interaction with matter, can escape from dense astrophysical environments that photons cannot, providing a unique glimpse into inner parts of stars and other non-transparent objects, while gravitational waves of various energetic bands carry complementary information about different processes in the Universe. 

\subsection{Electromagnetic radiation}
\label{sec:messengers:EM}~

With present-day astronomical instruments, photons from  space are detected in a very wide range of energies, from as low as $\sim 10^{-8}$~eV (radio waves, e.g.\ by the Low-Frequency Array, LOFAR) to as high as $\sim 10^{15}$~eV (very high energy gamma rays, e.g.\ by the Large High Altitude Air Shower Observatory, LHAASO). Existence of cosmic photons with even higher energies is guaranteed by the detection of ultra-high-energy cosmic rays, whose interactions with the cosmic microwave background should produce secondary photons. These photons, called cosmogenic ones, have not yet been observed. We note that photons with energies higher than MeV are detected with particle-physics methods, by means of recording individual events and not the intensity of the flux, so high-energy gamma-ray astronomy is quite different from more traditional observational one.

\subsection{Neutrinos}
\label{sec:messengers:nu}~

Astrophysical neutrinos are expected to span a similarly broad energy range, at least from $\sim 10^{-5}$~eV (relic neutrinos, born in the early Universe and presenting significant detection challenges because of extremely low interaction strength) to $\sim 10^{18}$~eV (cosmogenic neutrinos produced together with cosmogenic photons). However, low strength of neutrino interactions with matter and a large background of neutrinos from terrestrial processes allowed so far for the detection of astrophysical neutrinos in limited energy ranges only. These includes solar neutrinos (sub-MeV to a few MeV), neutrinos from a supernova explosion (few MeV to tens of MeV, see Sec.~\ref{sec:examples:SN1987A}) and a number of high-energy events from several TeV to several PeV. The interesting and rapidly developing field of the neutrino astronomy is covered by other lecturers here; for deeper reviews see e.g.\ Refs.~\cite{Spurio-book,ST-UFN,ST-UFN2023}.

\subsection{Gravitational waves}
\label{sec:messengers:GW}~

Gravitational waves are small perturbations of the spacetime metric, which are caused by the motion of masses and propagate across the Universe in a way similar to electromagnetic waves. There are certain analogies between gravitational and electromagnetic radiation, since both are described by a superposition of plane waves, propagating with the speed of light, and are caused by motion of sources (charges in the electromagnetic case). However, there are important differences. 

Electromagnetic waves are the solution to the Maxwell equations,
\begin{equation}
\partial_{\mu} F_{\mu\nu}= j_\nu,
\label{eq:Maxwell}
\end{equation}
where $F_{\mu\nu}=\partial_\mu A_\nu - \partial_\nu A_\mu$ is the electromagnetic stress tensor, linear in the electromagnetic field $A_\mu$, and $j_\mu$ is the source term describing charges and currents. Since the equation (\ref{eq:Maxwell}) is linear, application of the Fourier transform makes it easy to find that the solution is a combination of plane waves,
$$
A_\mu(\mathbf{x},t)= A_\mu^{(0)} \exp i(\mathbf{k} \mathbf{x}-\omega t).
$$

The gravitational field is described by the spacetime metric tensor $g_{\mu\nu}$, which satisfies the Einstein equations, 
\begin{equation}
G_{\mu\nu}=8\pi G T_{\mu\nu},
\label{eq:Einstein}
\end{equation}
where $G_{\mu\nu}$ is a \textit{nonlinear} function of $g_{\mu\nu}$, while $T_{\mu\nu}$ in the right-hand side is the energy-momentum tensor, describing the sources of the gravitational field: masses and their motion. Equation (\ref{eq:Einstein}) is therefore nonlinear, so unlike for Eq.~(\ref{eq:Maxwell}), finding its solutions is not straightforward. Gravitational waves are solutions to the linearized Einstein equations, valid for small perturbations $h_{\mu\nu}$ of the spacetime metric with respect to the metric $\eta_{\mu\nu}$ of the flat Minkowski space,
$$
g_{\mu\nu}=\eta_{\mu\nu}+h_{\mu\nu}, ~~ |h_{\mu\nu}| \ll 1.
$$
The linearized equation reads
\begin{equation}
\partial^\lambda \partial_\lambda h_{\mu\nu} (x) = -16 \pi G S_{\mu\nu}(x),
\label{eq:linearized}
\end{equation}
where 
$$
S_{\mu\nu} \equiv T_{\mu\nu} -\frac{1}{2} \eta_{\mu\nu} T^{\lambda~}_{~\lambda}.
$$
Equation (\ref{eq:linearized}), unlike the full Einstein equation (\ref{eq:Einstein}), allows for plane-wave solutions of the form
$$
h_{\mu\nu}(\mathbf{x},t)= h_{\mu\nu}^{(0)} \exp i(\mathbf{k} \mathbf{x}-\omega t).
$$

Detecting the electromagnetic and gravitational waves requires very distinct methodologies. The detectors of electromagnetic waves, like telescopes, cameras or human eyes, capture the intensity of the wave, which is related to the energy it carries and is proportional to the wave amplitude squared. On the other hand, the gravitational-wave detectors measure the actual deformations in spacetime caused by passing gravitational waves, that is the amplitude of the perturbation in the first power. The amplitude of a spherical wave propagating from a point source at the distance $D$ decreases as $1/D$, therefore the sensitivity of gravitational-wave detectors to distant sources decreases as $1/D$, compared to $1/D^2$ for conventional electromagnetic astronomy. 

Similarly to other messengers, cosmic gravitational waves also come at very different frequencies, which require very different detection methods. The gravitational-wave astronomy is still in its infancy, and only two ranges of frequencies have been tested so far: $\sim (10$~Hz $-10$~kHz) by terrestrial laser interferometers and $\sim$nHz by pulsar timing arrays.

\subsection{Cosmic rays}
\label{sec:messengers:CR}~

The term ``cosmic rays'', or, more precisely, ``cosmic-ray particles'', is used as a collective designation for all energetic particles in the Universe, except photons and neutrinos. It thus includes electrons and positrons, protons and nuclei, as well as other species, including hypothetical ones. While photons and neutrinos are stable and neutral, the majority of cosmic-ray particles are charged, hence deflected by cosmic magnetic fields. These deflections scale as $Z/E$ for particles with charge $Z$ and energy $E$. In the magnetic field of our Galaxy, even $E=10^{18}$~eV protons, $Z=1$, follow complicated paths, making it unrealistic to determine the source of any individual particle: the precise reconstruction of the trajectories would require an exact knowledge of the magnetic field's strength and configuration, details that are often not known accurately. Only at extreme energies of $E\sim 10^{20}$~eV, magnetic deflection of the particles decreases to several degrees, which is still far too much by astronomical standards, but opens up the possibility for speculations of their individual sources.

In addition, any deflection from a straight line means a longer trajectory, hence the time delay. For cosmic scales, even small deflections result in delays of thousands of years, thus making the time-domain astronomy with cosmic rays unfeasible even at the highest energies. 

An exceptional case within cosmic rays involves neutrons. Despite being unstable, high-energy neutrons can travel significant distances without decaying because of the relativistic time dilution which slows their decay from the perspective of an observer on Earth. For ultrarelativistic energies, $E \gtrsim 10^{18}$~eV, neutrons can carry information even from remote parts of the Milky Way. Clearly, the cosmic neutrons are much less abundant than their stable charged counterparts, protons and nuclei.

These complications suggest another approach to cosmic-ray astronomy, which proved to be useful also with other messengers (see Sec.~\ref{sec:diffuse}). It involves analyzing the diffuse overall flux of these particles. By studying the intensity of incoming cosmic rays, their distributions in energies and arrival directions, one can infer information both about their sources and about the medium through which they have traveled. This method does not rely on tracing the paths of individual particles but rather on the statistical analysis of their entire ensemble.

~

~

\section{Examples of individual multimessenger observations}
\label{sec:examples}
In this section, we discuss two key multimessenger observations, in both of which a short-duration event has been observed by means of two messengers. The examples represent important milestones in the development of the neutrino astronomy and the gravitational-wave research, respectively. 

\subsection{Supernova SN~1987A: neutrinos and electromagnetic waves}
\label{sec:examples:SN1987A}~

The first example is provided by the renowned supernova 1987A in the Large Magellanic Cloud (LMC), a dwarf satellite galaxy of the Milky Way. This event was detected on February 23, 1987, and generally confirmed existing theories about the mechanisms of core-collapse supernovae, which are thought to be the final evolutionary stages of massive stars. As these stars exhaust their nuclear fuel, their cores collapse under the force of gravity, releasing a significant amount of energy. Theoretically, most of this energy is expected to be emitted as neutrinos, which, due to their weak interaction with matter, escape from the core, see e.g.\ Refs.~\cite{ZasovPostnov,Janka}. This happens a few hours before the electromagnetic emission, coming from the outer expanding shell of the star, becomes visible as a supernova explosion. This process was evidenced when neutrino detectors registered the signal a few hours before the supernova became visible in electromagnetic observations.

The detectors involved were the Liquid Scintillator Detector (LSD) under Mont Blanc, the Baksan Underground Scintillator Telescope (BUST) in Kabardino-Balkaria, the Irvine-Michigan-Brookhaven (IMB) proton decay experiment in the United States, and Japan's Kamiokande~II. These detectors were designed to operate continuously and had a little, but not negligible chance to capture a rare core-collapse supernova in our Galaxy or in its immediate neighborhood. The SN~1987A progenitor star in LMC was  observed and studied before the explosion. At 2:52~UT on February 23, 1987, LSD, equipped with advanced software for its time, detected an unusual neutrino activity -- five neutrinos within a few seconds, an anomaly against the cosmic background. This detection led to an automatic alert disseminated to the scientific community, marking the beginning of a significant observation day.

Hours after the initial detection by the LSD, about 7:35~UT, three other detectors also registered neutrino bursts. The synchronization of the detectors' clocks was imprecise, lacking the exactness provided by modern GPS technology, leading to a few minutes of discrepancy in the recorded times of the neutrino detections. Nevertheless, astronomers posited that all detected neutrinos were emitted simultaneously, aligning within a ten-second window as per theoretical predictions. 

By 10:38~UT, the optical flare corresponding to the supernova was observed, illuminating the sky with a brightness comparable to the entire galaxy. Therefore, the neutrino flare was firmly associated with this event. The amount and energies of the neutrinos, the duration of the 7:35~UT neutrino burst, and the few-hours interval between the neutrino and visible flares, all confirmed -- up to available precision -- expectations from the mainstream core-collapse models. The statistics of only about two dozen events, as well as the approximate nature of the theoretical predictions, prevented one from a precise comparison. Clearly, the probability of a supernova explosion in our giant Milky Way was much higher than that in a dwarf, LMC, and in that case, the number of events would be much larger, because of the proximity. The detection of a double neutrino signal, which remains unexplained, continues to puzzle scientists. This anomaly represents one of the overlooked mysteries in astrophysics, not due to a lack of interest but rather because of the absence of a satisfactory explanation. More details on the observation of neutrinos from SN~1987A may be found e.g.\ in Ref.~\cite{Ryazhskaya}, see also Fig.~\ref{fig:SN}.
\begin{figure}[t]
\begin{center}
\includegraphics[width=134mm]{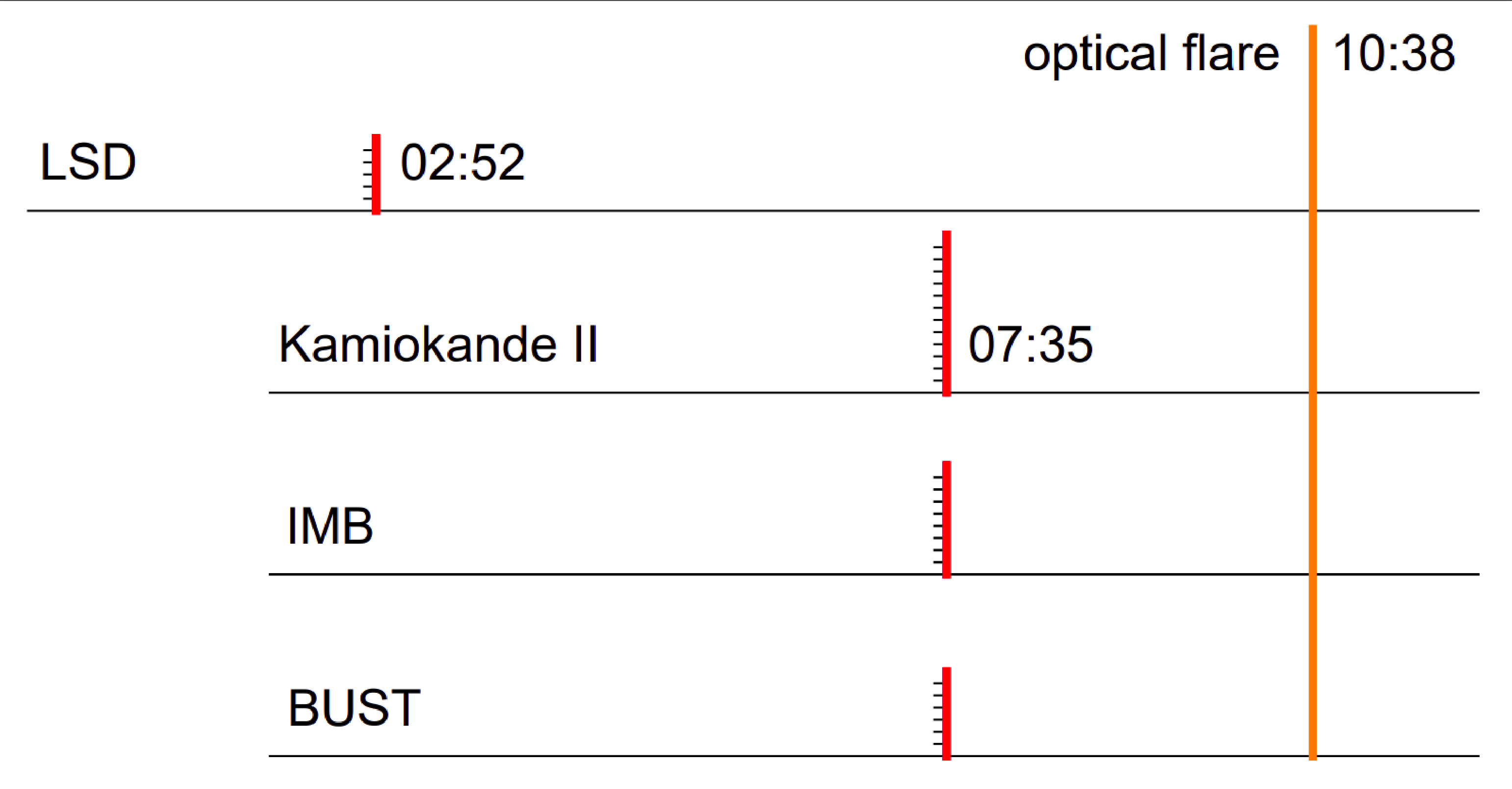}
\vspace{-3mm}
\caption{Timeline of observations of SN~1987A. For details and references, see Ref.~\cite{Ryazhskaya}.}
\end{center}
\labelf{fig:SN}
\vspace{-5mm}
\end{figure}

It is important to note that these observations also played a crucial role in constraining the properties of neutrino as a particle. These studies were based on the very fact of detection of neutrinos, arrived from $\sim 50$~kpc at the same day as the electromagnetic waves, and particularly constrain from above the neutrino mass, charge, and magnetic moment. Amusingly, the 2011 claim by the Oscillation Project with Emulsion-tRacking Apparatus (OPERA) collaboration, suggesting that neutrinos could travel faster than light, was in a serious contradiction with the SN~1987A observations, which implied the neutrino speed consistent with the speed of light to a precision of less than one part in ten billion. This contradiction highlighted potential issues within the OPERA findings, and the results about superluminal neutrinos have been subsequently retracted. In addition, the fact that the neutrino, and not other hypothetical light weakly interacting particles brought the gravitational energy from the supernova, allowed researchers to obtain constraints on properties of these particles never achieved in laboratory experiments, see e.g.\ Ref.~\cite{Raffelt}.

\subsection{Neutron-star merger GW~170817: electromagnetic and gravitational waves}
\label{sec:examples:GW170817}~

On August 17, 2017, another significant event was observed that exemplified the power of multi-messenger astronomy: the merger of two neutron stars detected through both gravitational waves and electromagnetic signals. This landmark observation underscored the advanced capabilities of current astronomical instruments and had profound implications for the understanding of fundamental physics.

The detection was not straightforward. Gravitational-wave telescopes, while revolutionary in their sensitivity to spacetime distortions, do not excel in pinpointing the origins of these waves with high precision by astronomical standards. In particular, the direction to the source of GW~170817 was described as a huge banana-shaped area in the sky. This uncertainty is partially compensated by the fact that the observation implies an estimate of the distance to the source even without its identification.

Simultaneously, spaceborne gamma-ray telescopes detected a burst within this same broad region, providing a rough timing and location that significantly narrowed down the search area. The synergy between these observations was crucial, as it allowed optical telescopes on Earth to quickly focus on a relatively small patch of the sky. There, astronomers discovered an optical flare, corresponding to a new light source that had not been present in images of the same part of the sky taken 20 days earlier. This flare was located in the candidate galaxy at the distance from the Earth which was in a good agreement with the gravitational-wave estimate. Altogether the observational picture confirmed the initial guess that this event is caused by merging two neutron stars, manifesting itself as a short gamma-ray burst, also known as a kilonova.  

The gravitational-wave measurement provided a detailed record of the change of frequency with time. For a binary system coalescence, the frequency of the emitted gravitational waves is one half of the orbital frequency, see e.g.\ Refs.~\cite{Capone,Weinberg}. These observations confirmed that as the neutron stars spiraled closer together and merged, the frequency of the emitted gravitational waves increased as predicted by general relativity. The decreasing orbital distance between the stars, inferred from the increasing frequency of the gravitational waves, was precisely measured in kilometers, despite the cosmic distance spanning a significant fraction of the size of the visible Universe.

This event not only validated the method of detecting gravitational waves but also played a crucial role in testing and refining theoretical models of gravity. The observed velocity of the gravitational waves, matching the speed of light, supported the predictions of general relativity and challenged some modified theories of gravity that predicted a difference in these velocities.

Simultaneous measurements of the amplitude and frequency of the gravitational waves from a merging relativistic binary enables one to reconstruct detailed information about the source, notably giving an independent measurement of the luminosity distance to the merger. Combining this measurement with electromagnetic observations, when available, opens up a possibility of determination of cosmological parameters, notably of the expansion rate of the Universe. This becomes particularly pertinent in addressing the so-called Hubble crisis, where differing measurements of the Hubble constant diverge by more than four standard deviations. The integration of data from gravitational waves with traditional electromagnetic observations could, over time and with accumulation of similar events, help resolve these discrepancies.

Impressive successes of the observations discussed in this section resulted in the widespread opinion that similar simultaneous observations of burst-like events constitute the essence of the multimessenger astronomy. We will immediately see below, however, that this is only partially so, and very important results may be obtained within a more general approach.

\section{Diffuse fluxes of different messengers}
\label{sec:diffuse}
Astrophysical high-energy neutrinos, with energies above several TeV and up to several PeV, have been conclusively detected by cubic-kilometer scale neutrino telescopes, IceCube in Antarctica and Baikal Gigaton Volume Detector (GVD) in Russia. Neutrinos with these high energies must be produced through non-thermal particle-physics processes. Indeed, the TeV temperatures required for the thermal production of high-energy neutrinos are millions times higher than e.g.\ those found in the core of a collapsing star, and are not present in the current Universe. 

The most plausible and nearly exclusive process identified for generating neutrinos above TeV involves hadronic collisions, specifically with the production of pi mesons. Upon decay, charged mesons $\pi^\pm$ produce neutrinos, while neutral $\pi^0$ mesons decay almost instantly into pairs of gamma rays. Astrophysical high-energy neutrinos serve as markers of relativistic hadrons and of non-thermal processes within the universe.

The propagation of high-energy photons in the Universe is limited due to the Universe's opacity caused by pair production on cosmic background radiation. The mean free path of photons depends on their energy and spans many orders of magnitude, ranging from the size of the Milky Way to the size of the visible Universe. The minimal path corresponds to the pair production on the Cosmic Microwave Background photons which are the most abundant. These photons are the targets for pair production by $\sim 100$~TeV photons, which therefore cannot reach us even from nearby galaxies, and, if observed, should be born in the Milky Way. High-energy photons from more distant sources undergo pair production; one of the resultant high-energy electrons then scatters off a cosmic background photon through inverse Compton scattering. The two processes, pair production and inverse Compton scattering, contribute to the development of the electromagnetic cascade, in which the average energy of photons decreases until their mean free path becomes very long. The energy of the initial photon is not lost but is instead re-emitted in softer photons, for which the Universe is transparent. These softer photons have energies in the (1--10)~GeV band, the spectral range observed e.g.\  by the Fermi Large Area Telescope (LAT), a gamma-ray telescope orbiting the Earth.

An important aspect of this phenomenon is that the directions of these secondary photons differ from that of the original gamma ray, because electrons and positrons are charged. As they traverse galactic and intergalactic magnetic fields, they are progressively deflected, increasingly losing the memory about their initial direction and contributing in the end to the diffuse isotropic gamma-ray flux, which Fermi LAT has measured.

Therefore, high-energy photons have to accompany the observed high-energy neutrinos at their birth, but their further fates may be different, depending whether the neutrinos and photons are produced in our Galaxy or in extragalactic sources. In the first case, one expects to detect energetic photons tracing the directions to the sites of the neutrino production. If the sources are extragalactic, then the electromagnetic cascades develop and contribute to the diffuse gamma-ray flux observed by Fermi LAT. This helps to disentangle the Galactic versus extragalactic origin of high-energy neutrinos. 

Spectra of high-energy neutrinos have been compiled from various analyses, mostly coming from the IceCube experiment. Notably, last year marked a significant milestone when the existence of these high-energy neutrinos was independently confirmed for the first time by a different experiment, Baikal-GVD. The Baikal-GVD spectrum agrees well with IceCube measurements. However, at energies $\sim 10$~TeV, yet unexplored by Baikal-GVD, IceCube results suggest the spectrum softer than one expects for extragalactic sources of nonthermal radiation. Assuming the standard production mechanism, one can estimate the total energy emitted in accompanying gamma rays, and obtain that it overshoots the energy in diffuse background photons measured by Fermi LAT. The flux of these Fermi-LAT photons is expected to have different contributions, including those unrelated to hadronic processes. The fact that the expected neutrino-related contribution by itself exceeds the measured values suggest that at least a part of 10-TeV neutrinos cannot be produced in extragalactic sources, pointing to the Galaxy as their origin.

In 2022--2023, these Galactic neutrinos were found in three independent analyses. Despite their methodological differences, these analyses roughly converged in their findings. Additionally, diffuse galactic gamma rays were detected by two other experiments slightly earlier, showing overall order-of-magnitude consistency across fluxes of different messengers, see Fig.~\ref{fig:Galactic}. 
\begin{figure}[t]
\begin{center}
\includegraphics[width=134mm]{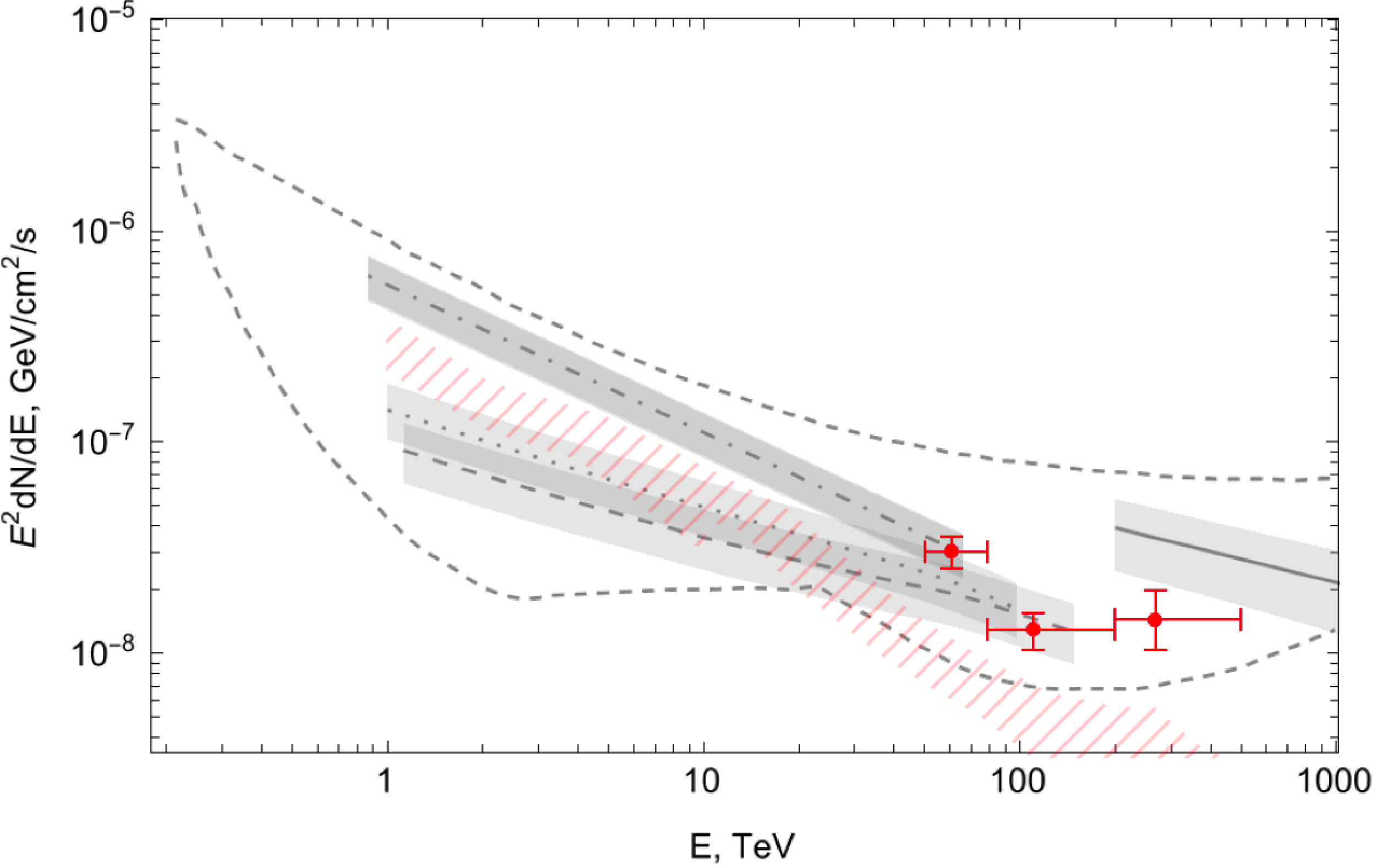}
\vspace{-3mm}
\caption{High-energy Galactic neutrino fluxes. Neutrinos from neutrino-telescope data: model-independent analysis of IceCube tracks (solid line at $E>200$~TeV; IceCube cascades under different model assumptions (dashed, dotted and dash-dotted lines); ANTARES (area delimeted by a short-dashed curve). Neutrinos as expected from gamma rays: Tibet-AS$\gamma$ (data points); LHAASO (hatched band). For further details and references, see Ref.~\cite{ST-UFN2023}.}
\end{center}
\labelf{fig:Galactic}
\vspace{-5mm}
\end{figure}

One of the searches for the Galactic neutrinos was performed in a purely model-independent way, which serendipitously revealed significant findings. Initially, the team was not specifically searching for Galactic neutrinos. However, upon mapping these neutrinos onto a sky map, they observed and quantified a significant concentration toward the Galactic plane. This revealed a puzzle which is yet to be fully interpreted. 

Indeed, the width of the visible Milky Way, corresponding to the thin Galactic disk, is a few angular degrees, that is a dozen of lunar visible angular diameters. Similar width of the Galactic emission in the sky was observed in energetic gamma rays by Fermi LAT. Contrasting sharply with these observations, the distribution of neutrinos appeared significantly wider. It extends up to 20$^\circ$ on either side of the Galactic plane, which greatly exceeds the width seen in the electromagnetic channel. Visually, this can be compared with the angular scale of the Big Dipper in the Ursa Major constellation, spanning about 15$^\circ$ in the sky. Other studies of the Galactic neutrino component \textit{assumed} the narrow disk from the very beginning.

The apparent broad distribution of neutrinos could potentially be explained by nearby astrophysical structures, such as the Local Bubble enriched with cosmic rays, particularly protons, and hosting regions of higher-than-average density of the interstellar gas. This bubble has actually been observed and described by astronomers, suggesting a localized overdensity that might be influencing the distribution patterns of Galactic neutrinos. Since this and similar structures are not accounted for by widely used models of Galactic cosmic rays, invoked in most of the neutrino studies, the width of the Milky Way in neutrinos became an important observable. It became evident that more refined models for cosmic ray production and propagation within galaxies are necessary to accurately represent these phenomena. This  exemplifies the essence of multi-messenger astronomy, which integrates observations from cosmic rays, neutrinos, and photons to derive comprehensive insights into cosmic phenomena. This approach not only helps in understanding individual cosmic events but also in piecing together the broader astrophysical context.

The discussion of neutrinos, photons, and cosmic rays' diffuse background sets the stage for another significant cosmic phenomenon: the diffuse gravi\-ta\-tio\-nal-wave background. It has been proposed in 1978 by Mikhail Sazhin, who passed away recently, that the gravitational waves with frequencies in the nanohertz band, potentially related to mergers of supermassive objects, may be detected by making use of the stable rotational periods of numerous pulsars scattered across the Galaxy, which serve as precise natural clocks. When a gravitational wave traverses Earth, it alters the space-time metrics, consequently changing the measured period of these pulsars. This shift in period occurs coherently across different directions, though not simultaneously, due to the nature of gravitational waves which do not impact all directions at once. This characteristic was a critical part of the hypothesis. In the summer of 2023, several research groups jointly reported the detection of this gravitational wave background, marking a significant milestone in astrophysics. These findings were based on observations spanning several decades, involving numerous pulsar readings. The angular correlation of these observations, measured versus the separation angle between pulsars, matched nicely the theoretical predictions, suggesting the successful detection of the gravitational wave background.

However, the spectra of these detected gravitational waves, if indeed caused by mergers of supermassive black holes, should be described by a particular expression, known as the Hellings-Downs spectrum (see Ref.\ \cite{HDcurveFAQ} for illuminating details). The observed spectrum demonstrated slight deviations from this predictions, which possibly imply new avenues for astrophysical and cosmological research beyond the established models.

\label{sec:Conclusion}
\section{Conclusions}

To summarize, astronomy has evolved into a multi-messenger discipline, extending beyond the traditional photonic realm. Multiple astrophysical messengers include neutrinos, gravitational waves, and cosmic rays, observed together with electromagnetic radiation. The integration of data from these diverse messengers enables a combined analysis that yields significant insights into astrophysical objects, media, and processes. This approach enhances the constraining power of observations, offering more precise and profound information than could be achieved through independent observations across different channels.

Importantly, this multimessenger approach is effective not only for analyzing specific cosmic events observed across two or more channels but also in studying the cumulative background signals from the full spectrum of physical sources. While no singular event has yet been observed simultaneously across three channels, the background data collected in this way have profound implications for fundamental physics.

One expects a bright future for the approaches described here, because many new instruments are just starting data taking. Almost surely, interesting observational results and discoveries will appear even before these lecture notes are published.

\label{sec:acknowledgement}
\section*{Acknowledgement}
We are indebted to the organizers of AYSS-2024 for their kind hospitality, as well as to all participants of the school for their important feedback.

\label{sec:funding}
\section*{Funding}
The work of ST on multimessenger astronomy is supported in the framework of the State project ``Science'' by the Ministry of Science and Higher Education of the Russian Federation under the contract 075-15-2024-541. 

\label{sec:conflict}
\section*{Conflict of interest}
The authors claim no conflict of interest.


\end{document}